\documentclass[pdflatex,sn-nature]{sn-jnl}

\usepackage{graphicx}%
\usepackage{multirow}%
\usepackage{amsmath,amssymb,amsfonts}%
\usepackage{amsthm}%
\usepackage{mathrsfs}%
\usepackage[title]{appendix}%
\usepackage{xcolor}%
\usepackage{textcomp}%
\usepackage{manyfoot}%
\usepackage{booktabs}%
\usepackage{algorithm}%
\usepackage{algorithmicx}%
\usepackage{algpseudocode}%
\usepackage{listings}%
\usepackage{siunitx}

\theoremstyle{thmstyleone}%
\theoremstyle{thmstyletwo}%

\theoremstyle{thmstylethree}%

\begin{document}
\title[Article Title]{Ultrafast all-optical quantum teleportation}


\author[1]{\fnm{Takumi} \sur{Suzuki}}

\author[1]{\fnm{Takaya} \sur{Hoshi}}

\author[1,2,3]{\fnm{Akito} \sur{Kawasaki}}

\author[1]{\fnm{Shotaro} \sur{Oki}}

\author[1]{\fnm{Konhi} \sur{Ichii}}

\author[1]{\fnm{Hironari} \sur{Nagayoshi}}

\author[1]{\fnm{Kazuma} \sur{Takahashi}}

\author[4]{\fnm{Takahiro} \sur{Kashiwazaki}}

\author[4]{\fnm{Taichi} \sur{Yamashima}}

\author[4]{\fnm{Asuka} \sur{Inoue}}

\author[4]{\fnm{Takeshi} \sur{Umeki}}

\author[1,2,3]{\fnm{Tatsuki} \sur{Sonoyama}}

\author[1,2,3]{\fnm{Kan} \sur{Takase}}

\author[1,2,3]{\fnm{Warit} \sur{Asavanant}}

\author[1,3]{\fnm{Mamoru} \sur{Endo}}

\author*[1,2,3]{\fnm{Akira} \sur{Furusawa}}\email{akiraf@ap.t.u-tokyo.ac.jp}

\affil*[1]{\orgdiv{Department of Applied Physics, School of Engineering}, \orgname{The University of Tokyo}, \orgaddress{\street{7-3-1 Hongo}, \city{Bunkyo}, \postcode{113-8656}, \state{Tokyo}, \country{Japan}}}

\affil[2]{\orgname{OptQC Corp.}, \orgaddress{\street{1-21-7 Nishi-Ikebukuro}, \city{Toshima}, \postcode{171-0021}, \state{Tokyo}, \country{Japan}}}

\affil[3]{\orgdiv{Optical Quantum Computing Research Team}, \orgname{RIKEN Center for Quantum Computing}, \orgaddress{\street{2-1 Hirosawa}, \city{Wako}, \postcode{351-0195}, \state{Saitama}, \country{Japan}}}

\affil[4]{\orgdiv{Device Technology Labs}, \orgname{NTT, Inc.}, \orgaddress{\street{3-1 Morinosato Wakamiya}, \city{Atsugi}, \postcode{243-0198}, \state{Kanagawa}, \country{Japan}}}

\maketitle

\textbf{Light's intrinsic carrier frequency of hundreds of terahertz theoretically enables information processing at terahertz clock rates \cite{Takeda2019-ze}. In optical quantum computing, continuous-variable quantum teleportation is the fundamental building block for deterministic logic operations \cite{Menicucci2006-ms,Ukai2010-eb,Asavanant2021-pr}. This protocol transfers unknown quantum states between nodes using quantum entanglement and real-time feedforward of measurement outcomes \cite{Vaidman1994-yv,Braunstein1998-fs,Furusawa1998-bd}. However, electrical feedforward bottlenecks currently restrict operational bandwidths to approximately 100 megahertz, preventing the exploitation of light's ultimate speed \cite{Lee2011-ox,Shiozawa2018-jf}. Here we show 1-terahertz-bandwidth all-optical quantum teleportation, completely bypassing this electronic limitation \cite{Ralph1999-xd}. By transferring Bell measurement outcomes optically, we successfully teleported vacuum states across the terahertz band and real-time random coherent wavepackets with a 42-picosecond temporal width. Evaluating the intrinsic state transfer quality, we achieved teleportation fidelities of $\mathcal{F} = 0.784$ for the broadband vacuum states and $\mathcal{F} = 0.770$ for the dynamic coherent wavepackets. Both results strictly surpass the classical limit of $\mathcal{F} = 0.5$ \cite{Braunstein2001-vc}, demonstrating genuine quantum teleportation at ultrafast speeds. Our results establish that optical quantum processing speeds are constrained solely by the nonlinear medium's 1-picosecond-scale response, rather than classical electrical interfaces. This methodology provides a cornerstone for terahertz-clock quantum computers capable of overcoming Moore's law, and paves the way for a high-capacity, telecom-compatible quantum internet \cite{Andersen2010-no}.}

\begin{figure}[htbp]
\centering
\includegraphics[width=\linewidth]{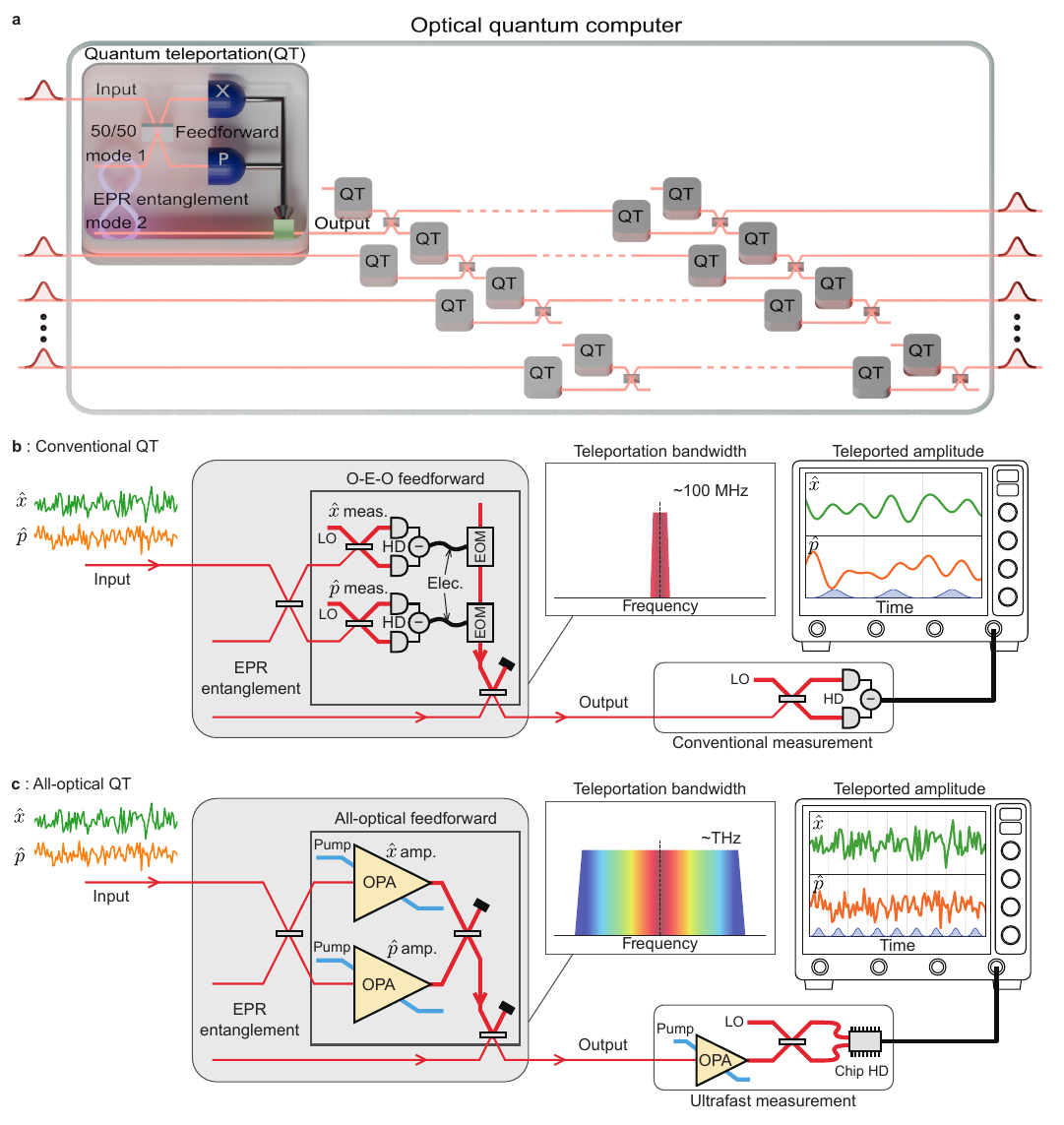} 

\caption{\textbf{All-optical quantum teleportation for high-speed measurement-based quantum computing.} 
\textbf{a}, Schematic of a multi-mode optical quantum computation architecture. Large-scale quantum circuits are executed through a sequence of quantum teleportations and beam-splitter operations. Within the quantum teleportation process, the input quadrature amplitudes $(\hat{x}_{\text{in}}, \hat{p}_{\text{in}})$ are transferred to the output modes via joint Bell measurements and subsequent feedforward operations. 
\textbf{b}, Conventional quantum teleportation utilizing an optical-to-electrical-to-optical (O-E-O) feedforward scheme. The operation bandwidth is fundamentally restricted to several hundred megahertz due to the bandwidth limits of the opto-electronic conversion components. 
\textbf{c}, All-optical quantum teleportation. By implementing the feedforward operation directly in the optical domain using OPAs, this architecture completely bypasses conventional electronic bottlenecks and enables terahertz-bandwidth processing of multi-mode quantum states.
EPR: Einstein-Podolsky-Rosen, HD: homodyne detector, EOM: electro-optic modulator, OPA: optical parametric amplifier, LO: local oscillator, pump: pump beam.}
\label{fig:concept}
\end{figure}

Modern supercomputers scale performance by massively parallelizing gigahertz (GHz) processors, driving energy consumption to unsustainable megawatt levels \cite{TOP500}. 
Quantum computers are anticipated as the ultimate saviors to break through this barrier, yet leading platforms---such as superconducting circuits and trapped ions---are restricted to sluggish kilohertz-to-megahertz clock rates. 
This profound speed deficit prevents them from outpacing classical supercomputers in actual wall-clock time for general-purpose tasks \cite{Babbush2021-ld}. 
Consequently, the realization of practical quantum advantage currently relies exclusively on exponential algorithmic speedups for specific applications, such as quantum chemistry.

Because light itself oscillates at hundreds of terahertz, optical quantum computing possesses the inherent potential to shatter this speed limit by operating at terahertz (THz) clock rates \cite{Takeda2019-ze}. 
Harnessing this vast carrier frequency as an information-processing bandwidth accelerates single-processor speeds by three orders of magnitude, establishing optical quantum computers as true general-purpose machines capable of executing both classical and quantum algorithms at unprecedented speeds. 
Crucially, this 1000-fold acceleration inherently resolves the catastrophic energy demands of massive parallelization, allowing the colossal hardware footprints and power consumption of modern supercomputing to be proportionally slashed. 
Thus, THz-clock optical processors offer the ultimate sustainable path for future global computing.

To realize this ultra-high-speed optical computer, the most promising architecture is measurement-based quantum computing (MBQC) \cite{Menicucci2006-ms,Ukai2010-eb}. 
In MBQC, large-scale quantum circuits are executed through a sequence of quantum teleportations \cite{Vaidman1994-yv,Braunstein1998-fs,Furusawa1998-bd} and beam-splitter operations (\textbf{Fig.~\ref{fig:concept}a}). 
Therefore, quantum teleportation is not merely a communication protocol; it is the fundamental active logic gate of an optical quantum computer, playing a role akin to that of the NAND gate in a classical CPU.

Continuous-variable (CV) quantum teleportation deterministically transfers quantum information by combining a quantum entangled state with a classical communication channel—specifically, a joint measurement followed by a feedforward operation \cite{Braunstein2005-lc}. 
While the entangled state perfectly preserves the quantum information, the operational speed of this logic gate is entirely dictated by the classical feedforward process. 
Consequently, achieving ultra-fast feedforward is the absolute prerequisite for realizing THz-clock optical quantum computing.

Realizing ultra-fast feedforward has remained the most formidable barrier since the first demonstration of deterministic CV quantum teleportation in 1998 \cite{Furusawa1998-bd}. In conventional optical-electrical-optical (O-E-O) approaches (\textbf{Fig.~\ref{fig:concept}b}), routing signals through the electrical domain inherently bottlenecks the operational bandwidth to the \qty{100}{\mega\hertz} regime due to slow opto-electronic conversions \cite{Lee2011-ox, Shiozawa2018-jf}, completely preventing the exploitation of light's ultimate THz potential and hindering the realization of high-speed, general-purpose optical quantum computers.

To bypass this electronic limitation, Ralph proposed ``all-optical quantum teleportation'' in 1999 \cite{Ralph1999-xd}, replacing O-E-O circuits with direct optical feedforward via phase-sensitive amplification (PSA). 
However, this elegant proposal remained experimentally elusive for over two decades. 
Traditional PSA relies on optical parametric oscillators (OPOs) \cite{Vahlbruch2016-lb}, where optical cavities strictly cap the bandwidth to the megahertz regime. 
Similarly, recent all-optical teleportation using atomic ensembles \cite{Liu2020-if} remains trapped in the megahertz order due to narrow atomic transition lines, rendering both approaches futile for THz-clock operation.

Recently, cavity-free waveguide optical parametric amplifiers (OPAs) have emerged as broadband alternatives \cite{Kashiwazaki2021-lw, Kashiwazaki2023-sx, Hirota2026-ew}, enabling high-speed PSA \cite{Inoue2023-ea, Nehra2022-fu, Kawasaki2024-kg, Kawasaki2025-pk, Yamashima2025-jd}.
Capitalizing on these advancements, we demonstrate deterministic all-optical CV quantum teleportation with a THz-class bandwidth. 
By entirely eliminating opto-electronic conversions, our setup directly processes joint measurement outcomes (\textbf{Fig.~\ref{fig:concept}c}); the weak quantum signals are optically amplified into robust macroscopic fields via high-gain PSA, allowing the deterministic application of feedforward displacements entirely within the optical domain. 

To prove picosecond-scale processing capabilities, we evaluate the system through two complementary measurements: a 1-THz ultra-broadband frequency-domain characterization of the teleported vacuum state, and a real-time time-domain measurement of 42-picosecond coherent-state wavepackets. 
These results resolve a quarter-century-old bottleneck, proving that the core logic gates of optical quantum computers can operate at THz clock rates.

\section*{Implementation of all-optical quantum teleportation}

\begin{figure}[htbp]
\centering
\includegraphics[width=\linewidth]{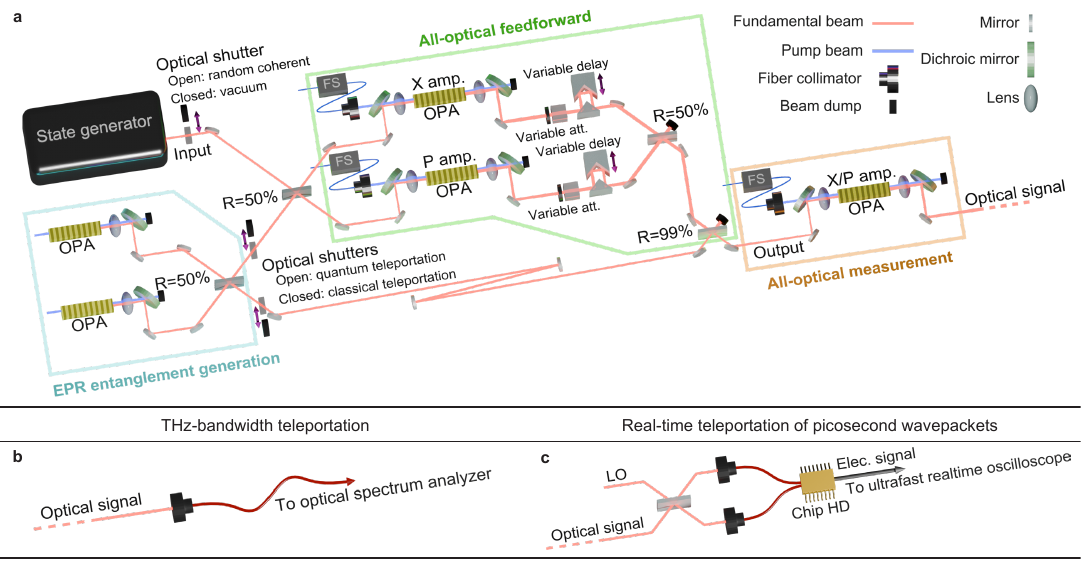} 

\caption{\textbf{Experimental setup for all-optical quantum teleportation.} 
\textbf{a}, All-optical quantum teleportation system. 
The setup comprises EPR entanglement generation, all-optical feedforward, and all-optical measurement, employing highly efficient PPLN waveguide optical parametric amplifiers (OPAs) throughout. 
Optical shutters switch the system between the quantum and classical teleportation regimes by selectively blocking the EPR output. 
\textbf{b}, Readout system. 
The evaluation setup supports two complementary measurement modes: frequency-domain characterization using an optical spectrum analyzer (OSA), and time-domain analysis using broadband homodyne detectors (HDs) coupled with an ultrafast real-time oscilloscope. 
FS: fiber stretcher, LO: local oscillator.}
\label{fig:setup}
\end{figure}

In the continuous-variable (CV) regime, quantum information is encoded in the quadrature amplitudes $\hat{x}$ and $\hat{p}$ \cite{Braunstein2005-lc}. Transferring an unknown input state $(\hat{x}_{\text{in}}, \hat{p}_{\text{in}})$ via measurement and feedforward operations inevitably contaminates the signal with quantum noise originating from ancillary modes (modes 1 and 2), yielding the output:
\begin{equation}\label{eq:output}
    \hat{x}_{\text{out}} = \hat{x}_{\text{in}} - \hat{x}_1 + \hat{x}_2, \quad \hat{p}_{\text{out}} = \hat{p}_{\text{in}} + \hat{p}_1 + \hat{p}_2.
\end{equation}
While relying on independent vacuum states (``classical teleportation'') leaves this noise uncancelled and strictly bounds the fidelity to the classical limit of $\mathcal{F} \le 0.5$ \cite{Braunstein2001-vc}, sharing an ideal Einstein-Podolsky-Rosen (EPR) entangled state completely circumvents this penalty. By exploiting the strong quantum correlations ($\hat{x}_1 - \hat{x}_2 \to 0$ and $\hat{p}_1 + \hat{p}_2 \to 0$), the noise terms perfectly cancel out, achieving perfect deterministic quantum teleportation ($\hat{x}_{\text{out}} \to \hat{x}_{\text{in}}$ and $\hat{p}_{\text{out}} \to \hat{p}_{\text{in}}$).

Figure~\ref{fig:setup} illustrates the experimental setup of our all-optical quantum teleportation system. 
We employ highly efficient periodically poled lithium niobate (PPLN) waveguide optical parametric amplifiers (OPAs) across all stages of the protocol: entanglement generation, all-optical feedforward, and quantum state measurement. 
The parametric processes in these OPAs are driven by a continuous-wave (CW) frequency-doubled pump beam at \qty{772.66}{\nano\meter}, corresponding to a fundamental telecommunication wavelength of \qty{1545.32}{\nano\meter}. 

As a benchmark for teleportation performance, we establish a classical teleportation baseline by blocking the EPR entanglement source with optical shutters. When utilizing these vacuum resources, the uncancelled noise terms in Eq.~\ref{eq:output} triple the output variance, yielding an output fluctuation level of precisely three times the shot noise limit (approximately \qty{+4.77}{\decibel}) \cite{Braunstein1998-fs}. Experimental observation of this theoretical value confirms the proper calibration of our all-optical feedforward.

Using this setup, we performed two complementary experiments to validate the system's ultrafast capabilities. 
First, we injected pure vacuum states to conduct terahertz-bandwidth optical spectrum measurements. 
However, an optical spectrum analyzer (OSA) yields only time-averaged statistical information due to its slow integration time. 
Because this stationary evaluation inherently lacks the dynamicity required for measurement-based quantum computing, frequency-domain analysis alone is insufficient.

As a second experiment to demonstrate true dynamic processing, we conducted real-time amplitude measurements in the time domain. We injected dynamic ``random coherent states'' with picosecond-scale temporal variations, mimicking the ultrafast computational data streams expected in a practical quantum processor. Capturing the real-time response to these ultrafast varying states is indispensable for proving the system's practical information processing capability.

\section*{Terahertz-bandwidth teleportation}
\begin{figure}[t]
\centering
\includegraphics[width=1\linewidth]{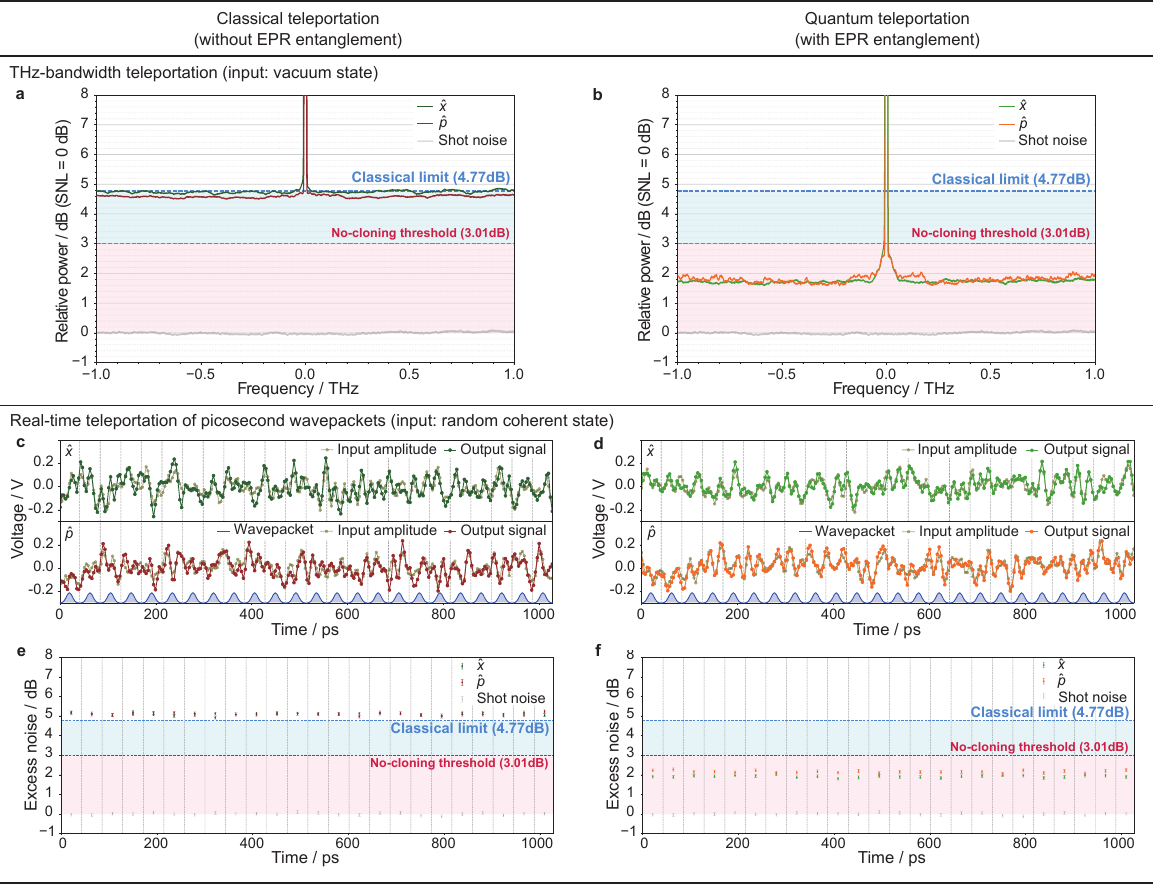}

\caption{\textbf{Experimental results of broadband and real-time quantum teleportation.} 
\textbf{a}, \textbf{b}, Optical spectra of the teleported output quadratures for a pure vacuum input state. Raw fluctuation power spectra (normalized to the shot-noise level at \qty{0}{\decibel}) of the $\hat{x}$- and $\hat{p}$-quadratures are shown for classical (\textbf{a}) and quantum (\textbf{b}) teleportation. The classical regime (\textbf{a}) shows raw variances of \qty{+4.74 \pm 0.06}{\decibel} ($\hat{x}$) and \qty{+4.58 \pm 0.06}{\decibel} ($\hat{p}$) averaged over the sidebands (\qtyrange{-1.0}{-0.2}{\tera\hertz} and \qtyrange{0.2}{1.0}{\tera\hertz}). In contrast, the quantum regime (\textbf{b}) demonstrates uniformly suppressed raw variances of \qty{+1.77 \pm 0.06}{\decibel} ($\hat{x}$) and \qty{+1.73 \pm 0.06}{\decibel} ($\hat{p}$). Accounting for the \qty{10}{\percent} final detection loss yields an intrinsic teleportation fidelity of $\mathcal{F} = 0.784 \pm 0.005$. 
\textbf{c, d}, Real-time voltage waveforms of the teleported output (solid lines) alongside the input random coherent amplitudes (light brown markers) for classical (\textbf{c}) and quantum (\textbf{d}) teleportation. Upper traces: $\hat{x}$-quadrature, lower traces: $\hat{p}$-quadrature. Blue shaded areas represent the \qty{42}{\pico\second} temporal wavepackets.
\textbf{e}, \textbf{f}, Time-resolved intrinsic variances of the quantum fluctuations. Similarly accounting for the \qty{10}{\percent} detection loss, the quantum setup (\textbf{f}) maintains uniformly suppressed variances, corresponding to an intrinsic teleportation fidelity of $\mathcal{F} = 0.770 \pm 0.006$.}
\label{fig:results}
\end{figure}

Figures~\ref{fig:results}a and \ref{fig:results}b present the frequency-domain performance evaluation of our quantum teleportation system. By injecting a pure vacuum state into the signal port, we measured the power spectra of the teleported output states across a broad bandwidth spanning from \qty{-1.0}{\tera\hertz} to \qty{+1.0}{\tera\hertz} relative to the optical carrier frequency. 

First, to establish a classical teleportation baseline, we evaluated the system without entanglement. In our architecture, the optical loss in the all-optical feedforward increases the output variance, whereas the detection loss in the all-optical measurement attenuates it. When these two losses are identical, they cancel each other out, and the raw variance coincides with the \qty{+4.77}{\decibel} theoretical limit. When averaged over the sideband frequencies from \qty{-1.0}{\tera\hertz} to \qty{-0.2}{\tera\hertz} and \qty{0.2}{\tera\hertz} to \qty{1.0}{\tera\hertz}, the observed raw variances were indeed \qty{+4.74 \pm 0.06}{\decibel} ($\hat{x}$) and \qty{+4.58 \pm 0.06}{\decibel} ($\hat{p}$) (\textbf{Fig.~\ref{fig:results}a}). Furthermore, by accounting for the estimated \qty{10}{\percent} detection loss in the all-optical measurement, the intrinsic variances of the teleported output evaluate to \qty{+5.05 \pm 0.06}{\decibel} and \qty{+4.88 \pm 0.06}{\decibel}. This proximity to the ideal limit unambiguously confirms the successful implementation of accurate feedforward operations across the exceptionally broad \qty{1}{\tera\hertz} sideband bandwidth.

Next, introducing the EPR entanglement uniformly suppressed the quantum fluctuations across the entire measured bandwidth (\textbf{Fig.~\ref{fig:results}b}). The observed raw variances dropped to \qty{+1.77 \pm 0.06}{\decibel} ($\hat{x}$) and \qty{+1.73 \pm 0.06}{\decibel} ($\hat{p}$). By accounting for the final \qty{10}{\percent} detection loss, we reconstructed the intrinsic excess noise at the teleporter's output to be \qty{+1.93 \pm 0.06}{\decibel} and \qty{+1.88 \pm 0.06}{\decibel}, respectively.

Finally, evaluating the overall state transfer quality, we achieved a raw teleportation fidelity of $\mathcal{F} = 0.801 \pm 0.004$ ($\mathcal{F} = 0.784 \pm 0.005$ based on the inferred intrinsic variances). These results conclusively shatter the classical limit of $\mathcal{F} = 0.5$ (corresponding to \qty{+4.77}{\decibel} of excess noise), providing definitive proof of successful quantum teleportation. Furthermore, the system strictly surpasses the more stringent quantum no-cloning limit of $\mathcal{F} = 2/3 \approx 0.667$ \cite{Grosshans2001-ow} (corresponding to \qty{+3.01}{\decibel} of excess noise) across all evaluated scenarios. Note that all specified error bars denote one standard deviation, primarily originating from a slight fluctuation ($\sigma = \qty{0.06}{\decibel}$) in the measurement OPA gain. These findings establish that our all-optical approach is fundamentally capable of high-fidelity quantum operations at terahertz-class clock frequencies.

\section*{Real-time teleportation of picosecond wavepackets}

Figures~\ref{fig:results}c and \ref{fig:results}d display the real-time voltage waveforms obtained when random coherent states were injected into the system, serving as a primary highlight of our time-domain evaluation. The traces display the input signal amplitudes (light brown) alongside the teleported output signals for both classical teleportation (\textbf{Fig.~\ref{fig:results}c}; dark green for $\hat{x}$, brown for $\hat{p}$) and quantum teleportation (\textbf{Fig.~\ref{fig:results}d}; green for $\hat{x}$, orange for $\hat{p}$). Across all these signals, the real-time trajectories exhibit a clear correlation with the input despite the picosecond-scale rapid transitions, visually demonstrating the ultra-fast response of the all-optical feedforward.

To quantify this dynamic performance, we analyzed the statistical distribution of the quadrature amplitudes for the time-series data by extracting consecutive independent temporal modes (\textbf{Figs.~\ref{fig:results}e} and \textbf{\ref{fig:results}f}). To guarantee that adjacent wavepackets are statistically uncorrelated, each mode is defined by a full integration window of \qty{42}{\pico\second}---the specific duration required for the autocorrelation function of the input coherent amplitude to decay to zero. We then accounted for the \qty{10}{\percent} measurement inefficiency to rigorously reconstruct the intrinsic state transfer quality immediately at the teleporter's output. Establishing the classical baseline without entanglement, the inferred intrinsic variances were evaluated as \qty{+5.12 \pm 0.08}{\decibel} ($\hat{x}$) and \qty{+5.09 \pm 0.08}{\decibel} ($\hat{p}$) (\textbf{Fig.~\ref{fig:results}e}). Next, upon introducing the EPR entanglement, the intrinsic variances of the quantum fluctuations were successfully suppressed to \qty{+1.92 \pm 0.08}{\decibel} ($\hat{x}$) and \qty{+2.16 \pm 0.08}{\decibel} ($\hat{p}$) (\textbf{Fig.~\ref{fig:results}f}).

By accounting for measurement inefficiencies to evaluate the intrinsic state transfer quality, the teleportation fidelity reaches $\mathcal{F}_{\text{int}} = 0.770 \pm 0.006$ (evaluated as $\bar{\mathcal{F}}_{\text{raw}} = 0.776 \pm 0.006$ when analyzed directly from the raw signals). Crucially, both the intrinsic and raw performances strictly surpass the classical limit of $\mathcal{F} = 0.5$ as well as the quantum no-cloning threshold of $\mathcal{F} = 2/3 \approx 0.667$ \cite{Grosshans2001-ow}, successfully demonstrating genuine quantum teleportation. Note that all specified error bars denote one standard deviation, derived by comprehensively accounting for both statistical uncertainty (\qty{0.06}{\decibel}) and the influence of measurement OPA gain fluctuations (\qty{0.06}{\decibel}).

\section*{Discussion}

We have demonstrated the complete circumvention of the opto-electronic conversion bottleneck in continuous-variable quantum teleportation through an all-optical feedforward architecture \cite{Yamashima2025-jd}, achieving terahertz-bandwidth operation---a 10,000-fold improvement over conventional electronic methods limited to the \qty{100}{\mega\hertz} regime \cite{Lee2011-ox,Shiozawa2018-jf}. 
Limited only by passive optical losses, the intrinsic teleportation fidelities for both broadband vacuum ($\mathcal{F} = 0.784$) and dynamic coherent wavepackets ($\mathcal{F} = 0.770$) definitively surpass the classical limit of $\mathcal{F} = 0.5$ \cite{Braunstein2001-vc}. 
Furthermore, they even exceed the quantum no-cloning threshold of $\mathcal{F} = 2/3$ \cite{Grosshans2001-ow}, achieving fidelities high enough to support the non-Gaussian state teleportation indispensable for fault-tolerant quantum computing. 
Crucially, while our real-time observation was bounded to a \qty{42}{\pico\second} window by the readout electronics, the confirmed \qty{1}{\tera\hertz} frequency-domain bandwidth guarantees that the intrinsic optical processing operates at the \qty{1}{\pico\second} level. 
Notably, our architecture inherently circumvents this readout bottleneck. Because fragile quantum states are optically amplified into macroscopic classical fields prior to detection, we can directly incorporate established ultrafast classical photonics techniques---such as wavelength-division multiplexing and electro-optic sampling---for even faster information extraction.

This terahertz-level clock rate establishes a decisive hardware advantage over other leading quantum computing platforms. 
While matter-based architectures---such as superconducting circuits and trapped ions---are fundamentally restricted to the kilohertz-to-megahertz regime, our architecture fully unleashes the immense bandwidth of the optical carrier. 
This definitively elevates optical quantum processors to the undisputed fastest physical platform, operating at clock frequencies tens of thousands of times higher than existing architectures. 

Beyond raw speed, this terahertz clock rate fundamentally transforms the scalability of time-domain multiplexed measurement-based quantum computing (MBQC) \cite{Menicucci2006-ms,Asavanant2019-eg,Larsen2019-eu}. 
In such architectures, the number of multiplexed qu-modes is directly dictated by the temporal duration of the quantum wavepackets accommodated within a fixed optical fiber loop. 
By shortening this duration to the \qty{1}{\pico\second} level, our methodology enables a dramatic increase in the number of qu-modes without expanding the physical footprint of the system. 
Consequently, realizing a massive-scale processor with over one million qu-modes now emerges as a highly realistic experimental parameter.

While classical supercomputers rely on the brute-force parallelization of gigahertz processors, inevitably demanding megawatt-level energy consumption, our optical platform directly overcomes these physical limits of modern supercomputing through the synergy of this terahertz clock rate and million-mode scalability. 
By combining the inherent exponential speedup of quantum algorithms with this immense terahertz hardware acceleration, we establish a realistic pathway toward general-purpose high-speed computation capable of outperforming classical supercomputers in absolute wall-clock time.

Beyond standalone computation, the applications of this broadband optical teleportation technology extend directly to existing optical fiber networks and broader optical communication infrastructures. 
Furthermore, its terahertz bandwidth perfectly aligns with anticipated sixth-generation (6G) networks, establishing a core enabling technology for ultrafast quantum repeaters that will drive high-capacity quantum communication and the global quantum internet.

In conclusion, our all-optical quantum teleportation endows quantum processors with an unprecedented dual advantage: the exponential speedup inherent to quantum algorithms combined with the immense linear acceleration of terahertz clocks. 
Ultimately, this work establishes a solid foundation for the realization of a true all-photonic quantum supercomputer.

\bibliography{Paperpile}

\newpage
\section*{Methods}

\subsection*{Master laser and nonlinear waveguide devices}
The master light source was a narrow-linewidth continuous-wave (CW) fiber laser system (Koheras HARMONIK HP H78, NKT Photonics) providing approximately \qty{7}{\watt} of fundamental power (\qty{1545.32}{\nano\meter}) and \qty{8}{\watt} of second-harmonic power (\qty{772.66}{\nano\meter}). The entire experimental setup was constructed as a hybrid system utilizing both fiber-coupled components and free-space optics. For entanglement generation, all-optical feedforward, and all-optical measurement, we employed \qty{45}{\milli\meter} ZnO-doped periodically poled lithium niobate (PPLN) ridge waveguides \cite{Kashiwazaki2021-lw, Kashiwazaki2023-sx}. Detailed parameters regarding mode-matching efficiencies, optical loss budgets, and feedforward gain calibration methods are provided in the Supplementary Information.

\subsection*{Phase stabilization and sample-and-hold data acquisition}
Active phase stabilization of the complex hybrid interferometer network was achieved using auxiliary probe beams and lock-in detection. For real-time time-domain measurements, the continuous presence of these probe beams would occupy a portion of the limited dynamic range of our high-speed oscilloscope (effective number of bits, ENOB $\approx \qty{5}{bits}$), severely degrading the vertical resolution available for the microscopic quantum fluctuations. To circumvent this, we adopted a sample-and-hold control scheme in real-time measurements. The measurement sequence was time-division multiplexed: the system was actively stabilized during the locking interval, and the feedback voltages were held constant during the measurement interval while the probe beams were completely blocked, enabling artifact-free real-time observation.

\subsection*{Effective efficiency in all-optical processing}
In our all-optical processing scheme, the physical optical loss experienced inside the waveguide does not directly correspond to the loss of quantum information. Because phase-sensitive amplification (PSA) and loss occur simultaneously along the distributed PPLN waveguide, the impact of internal loss is effectively suppressed by the parametric gain \cite{Yamashima2025-jd}. Specifically, the feedforward operation and the pre-amplification before measurement operated with parametric gains of \qty{30}{\decibel} and \qty{25}{\decibel}, respectively. This fundamental mechanism highly improves their effective efficiencies to approximately \qty{98.8}{\percent} and \qty{98.6}{\percent}. Furthermore, this optical pre-amplification strongly mitigates the measurement loss originating from the finite intrinsic quantum efficiency (\qty{30}{\percent}) of the high-speed homodyne detector, maintaining an exceptionally high overall homodyne efficiency of approximately \qty{97.8}{\percent} \cite{Inoue2023-ea, Kawasaki2024-kg, Kawasaki2025-pk}. Full analytical derivations are provided in the Supplementary Information.

\subsection*{Broadband frequency-domain evaluation}
Frequency-domain measurements were performed using an optical spectrum analyzer (OSA). Under strong PSA gain conditions, the optical power spectral density measured by the OSA is directly proportional to the variance of the quadrature amplitude subjected to PSA \cite{Shaked2018-zj, Takanashi2020-hb, Kashiwazaki2021-lw, Nehra2022-fu, Yamashima2025-jd}. Teleportation of a broadband vacuum state was evaluated over a terahertz-class bandwidth. To isolate the optical signal of the system output from technical low-frequency excess noise (e.g., master laser phase noise), the signal intensity levels and the resulting teleportation fidelity $\mathcal{F}$ were calculated by averaging the power spectra over the flat regions spanning \qtyrange{-1.0}{-0.2}{\tera\hertz} and \qtyrange{0.2}{1.0}{\tera\hertz}.

\subsection*{Real-time time-domain evaluation}
For time-domain performance evaluation, we utilized broadband optical emission from a superluminescent diode (SLD) to prepare random coherent states. The emission was spectrally filtered to a bandwidth of approximately \qty{110}{\giga\hertz} matching the master laser carrier, corresponding to a baseband electrical bandwidth of \qty{55}{\giga\hertz}. The macroscopic fluctuations were continuously monitored by an independent dual-homodyne setup before being heavily attenuated by \qty{25}{\decibel} to project them into the quantum regime. 
The teleported output was pre-amplified all-optically and measured using a \qty{70}{\giga\hertz}-bandwidth balanced photodetector and a real-time oscilloscope (\qty{110}{\giga\hertz} analog bandwidth, \qty{256}{GSa/s} sampling rate). Synchronized with the sample-and-hold interval, we acquired 128 individual measurement traces. To define the single-mode quadrature amplitude $\hat{x}_k$ for the $k$-th wavepacket, we performed a digital weighted integration using a \qty{42}{\pico\second} Gaussian window over the sampled data points. Mathematical formulations for the general state transfer fidelity are detailed in the Supplementary Information.

\subsection*{Theoretical noise model}
To quantitatively validate the measured excess noise levels, we constructed a noise build-up model incorporating the key experimental parameters: the effective squeezing noise of the EPR resource $N_{\text{sq}} = 0.178$ (corresponding to \qty{7.5}{\decibel} squeezing), the readout efficiency of the Bell measurement $\eta_{\text{Bell}} = 0.900$, and the final readout efficiency $\eta_{\text{meas}} = 0.900$. The normalized output noise variance $N_{\text{out}}$ for vacuum teleportation is calculated as:
\begin{equation}
    N_{\text{out}} = \eta_{\text{meas}} \left( 1 + 2N_{\text{sq}} + \frac{2(1-\eta_{\text{Bell}})}{\eta_{\text{Bell}}} \right) + (1 - \eta_{\text{meas}}).
\end{equation}
This model predicts an output noise of \qty{1.82}{\decibel} ($N_{\text{out}} \approx 1.52$) for quantum teleportation and \qty{4.77}{\decibel} ($N_{\text{out}} = 3.00$) for classical teleportation without entanglement ($N_{\text{sq}} = 1$). These theoretical values are in good agreement with our experimentally observed raw excess noise levels, demonstrating that this model accurately captures the primary loss and noise dynamics of our continuous-variable teleportation system.

\subsection*{Use of AI tools}
During the preparation of this manuscript, the authors used Gemini 3.1 Pro (Google) and Claude Opus 4.6 (Anthropic) to translate initial drafts, refine the English writing, and improve the overall readability of the text. After using these tools, the authors thoroughly reviewed and edited the content as needed, and take full responsibility for the final publication.

\section*{Data availability}
The data that support the findings of this study are available from the corresponding author upon reasonable request.

\section*{Acknowledgements}
The authors acknowledge support from UTokyo Foundation and donations from Nichia Corporation of Japan.
This work was supported by Japan Science and Technology Agency (JST) Moonshot R\&D Grant Nos. JPMJMS2064 and JPMJMS256I.
A part of this work was also performed for Council for Science, Technology and Innovation (CSTI), Cross-ministerial Strategic Innovation Promotion Program (SIP), ``Promoting Application of Advanced Quantum Technologies to Social Challenges'' (Project management agency: QST).
T. Suzuki, A. Kawasaki, H. Nagayoshi, K. Takahashi, and T. Sonoyama acknowledge financial support from The Forefront Physics and Mathematics Program to Drive Transformation (FoPM), a World-leading Innovative Graduate Study (WINGS) Program.
T. Suzuki (No. 25KJ0768), A. Kawasaki (No. 24KJ0644), H. Nagayoshi (No. 24KJ0745), K. Takahashi (No. 25KJ1172), and T. Sonoyama (No. 23KJ0518) acknowledge support from Japan Society for the Promotion of Science (JSPS) Research Fellowship for Young Scientists.
A. Kawasaki also acknowledges support from Leadership Development Program for Ph.D. (LDPP).
W.A. acknowledges funding from JSPS KAKENHI (No. 23K13040).
K. Takase acknowledges funding from JSPS KAKENHI (No. 23K13038, 22K20351).
M.E. acknowledges funding from JST (JPMJPR2254) and JSPS KAKENHI (No. 24K01374).
The authors also thank NTT Device Technology Labs for providing the PPLN waveguides used in this study.

\section*{Author contributions}
T. Suzuki conceived and designed the experiments and led the data evaluation and analysis. T. Suzuki, T.H., A.K., S.O., K.I. and K. Takahashi developed the experimental system, with T. Suzuki leading the implementation. T. Kashiwazaki, T.Y., A.I. and T.U. prepared the PPLN crystals. H.N. contributed to the evaluation and analysis of the experimental results. T. Sonoyama, K. Takase, W.A., M.E. and A.F. supervised the project. All authors discussed the results and contributed to the manuscript.

\section*{Competing interests}
The authors declare no competing interests.

\end{document}